\documentclass{mhd}     
\usepackage{amssymb,amsmath,subfigure,graphicx}
\usepackage{eurosym,color}

\newcommand{\frage}[1]{#1}
\newcommand{\bi}{ \boldsymbol}

\mhdhead{0}{0}{1}	

\title{Metal pad roll instability in liquid metal batteries}	

\author{N.~Weber\inst{1}, P. Beckstein\inst{1}, V. Galindo\inst{1}, W. Herreman\inst{2},\\ C. Nore\inst{2}, F. Stefani\inst{1} T. Weier\inst{1}} 	
\institute{Full postal address of the author's institute} 

\institute{Helmholtz-Zentrum Dresden - Rossendorf, Bautzner
  Landstr. 400, 01328 Dresden, Germany
\and Laboratoire d'Informatique pour la M\'ecanique et
les Sciences de l'Ing\'enieur, CNRS UPR 3251, b\^at. 508, 91405 Orsay
CEDEX and Universit\'e Paris-Sud 11, France}

\begin{document}

\maketitle
\begin{abstract}%
The increasing deployment of renewable energies requires three
fundamental changes to the electric grid: more transmission lines, a
flexibilisation of the demand and grid scale energy storage. Liquid
metal batteries (LMBs) are considered these days as a promising means
of stationary energy storage. Built as a stable density stratification of
two liquid metals separated by a liquid salt, LMBs have three main advantages:
a low price, a long life-time and extremely high current densities. In order
to be cheap, LMBs have to be built large. However, battery currents in the order
of kilo-amperes may lead to magnetohydrodynamic (MHD) instabilities, which -- in the worst case -- may
short-circuit the thin electrolyte layer. The metal pad roll instability, as known
from aluminium reduction cells, is considered as one of the most dangerous phenomena
for LMBs. We develop a numerical model, combining fluid- and
electrodynamics with the volume-of-fluid method, to simulate this
instability in cylindrical LMBs. We explain the instability mechanism
similar to that in aluminium reduction cells and give some first results,
including growth rates and oscillation periods of the
instability\footnote{This article is an extended version of a conference paper
of the proceedings of the 10th PAMIR conference \cite{Weber2016b}.}.
\end{abstract}

\section{Motivation}
\begin{figure}[b!]
\centering
\subfigure[]{\includegraphics[height=4.5cm]{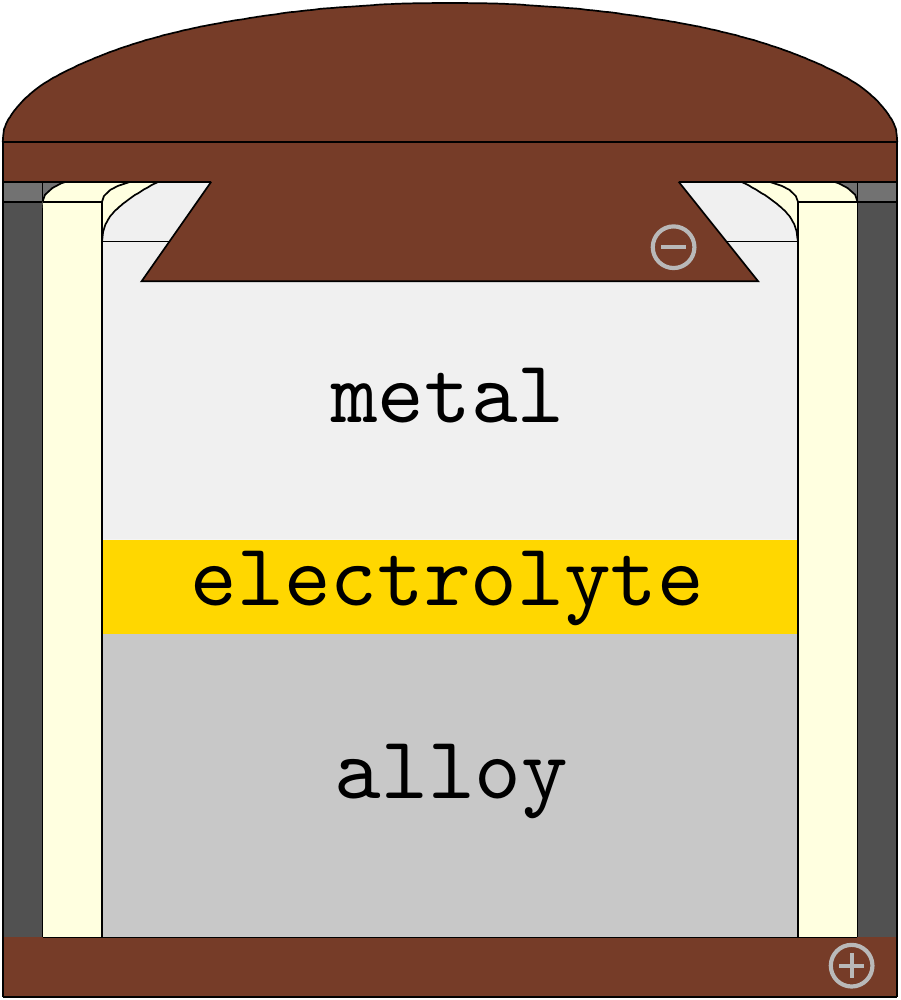}}\hspace{4cm}
\subfigure[]{\includegraphics[height=4.5cm]{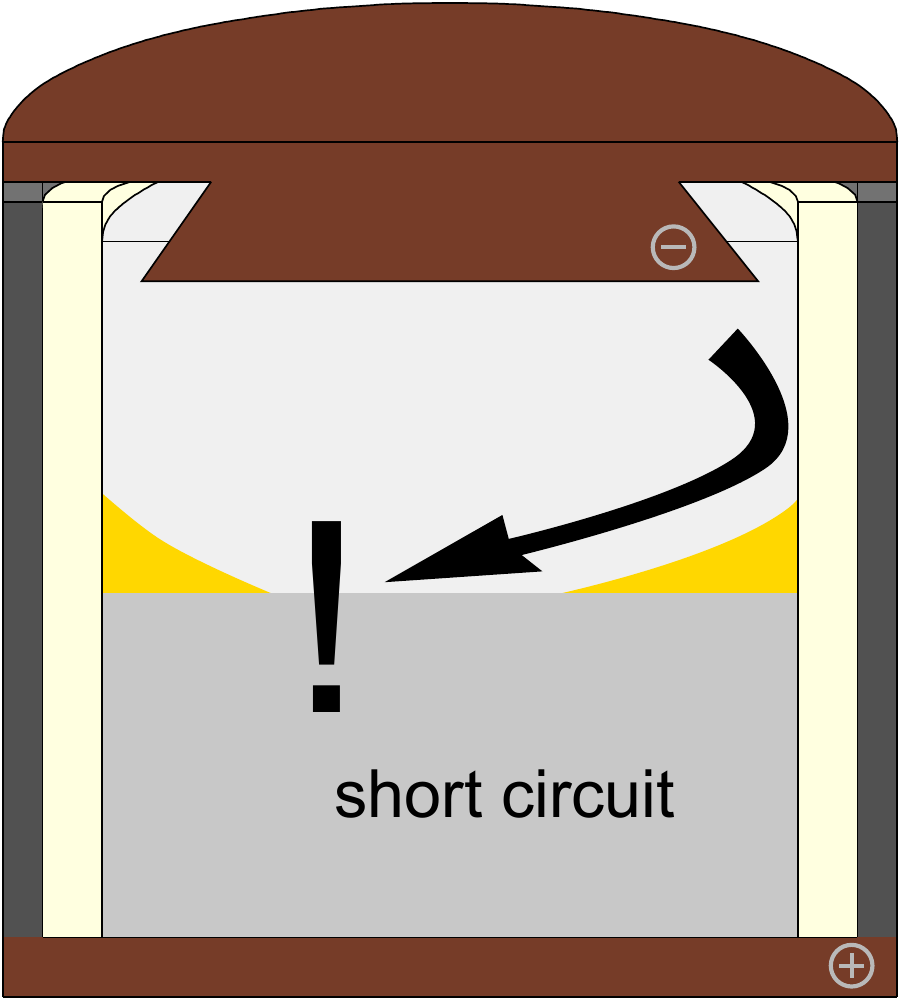}}
\caption{Scheme of a liquid metal battery with typical inventory (a),
and short circuit due to a strong fluid flow in the upper metal compartment (b).}
\label{f:lmb}
\end{figure}
According to prognoses of the International Energy Agency, the worldwide
energy demand will grow from the year 2011 to 2035 by two thirds. In the same period,
the share of renewable energies is predicted to rise from 20 to 31\,\%
\cite{IEA2013}. These renewable energies are highly fluctuating; in order to
stabilise voltage and frequency in the electric grid, new transmission
lines must be built and the demand should be flexibilised as much as
possible. Furthermore, stationary energy storage on a large scale will
become mandatory. Typical requirements for such storage
devices are a low price ($<0.1$\euro/kWh), a long life-time ($>7000$ cycles),
fast response times, and safety \cite{Huggins2016,Noorden2014,Bradwell2011,Kim2013b}.
Liquid metal batteries (LMBs) promise to fulfil these requirements \cite{Kim2013b};
they are high temperature batteries as e.g. sodium sulphur and ZEBRA
(NaNiCl$_2$) cells \cite{Hueso2013,Liu2013}. LMBs consist of a stable
density stratification of two liquid metals, separated by a molten salt layer
(Fig. \ref{f:lmb}a) \cite{Cairns1967, Crouthamel1967}. During
discharge, the upper metal will donate electrons. The ion will cross
the electrolyte layer and alloy with the bottom metal. If no corrosion
occurs, the reaction should be fully reversible. 

First liquid metal cells were operated in the 1960 by General
Motors Corporation, Atomics International and Argonne National
Laboratory mostly for the conversion of thermal to electric energy.
Cairns \cite{Cairns1969b} gives a short overview
about these works; Crouthamel \cite{Crouthamel1967}, Cairns
\cite{Cairns1967}, Swinkels \cite{Swinkels1971} and Chum
\cite{Chum1980,Chum1981} provide very detailed information. After some
calm decades, liquid metal batteries were revived by the group of
Donald Sadoway \cite{Bradwell2011, Kim2013b} as grid scale energy
storage devices. Their main argument for the deployment of LMBs is the low price of these cells.

As the active materials of LMBs
are predicted to represent only 25\,\% of the cell cost, LMBs have to be
built large, in order to be cheap. However, strong currents may
lead to different instabilities. In the worst case, a strong fluid flow may wipe away
the electrolyte layer and short-circuit the battery (Fig. \ref{f:lmb}b). Depending on the
geometry of the current collectors, electro-vortex flow can be dangerous
for batteries of a diameter larger than several decimetres \cite{Weber2014b}. The
current-driven Tayler instability may appear, too, and lead to a short
circuit in very large cells (diameter $>1$\,m and aspect ratio of the same order) \cite{Weber2014,
Herreman2015}. Further, the high resistance of the salt layer will lead to
internal heat generation and thermal convection in the upper metal as
well as the electrolyte layer, although the resulting flow is not expected to
endanger the operation of small and medium sized cells
\cite{Shen2015}. Thermal convection was studied in addition with the
intention to increase mass transfer in the lower metal and improve the
efficiency of the cell \cite{Kelley2014,Beltran2016}. Concentration
polarisation and even the formation of solid intermetallic phases are
indeed important issues because this is often observed experimentally
\cite{Cairns1967,Kim2013a,Heredy1967,Agruss1967,Foster1967b,Vogel1967,Ouchi2014}.

Long wave interface instabilities, also known as metal pad rolling or
sloshing, may lead to a short-circuit even in small cells
\cite{Zikanov2015}. These instabilities are already known from aluminium smelters,
where they limit the maximum electrolysis current \cite{Sele1977,
  Bojarevics1994, Davidson2001, Molokov2011}. Here, we use an
integro-differential MHD model \cite{Weber2013} as already used to
simulate the Tayler instability and electro-vortex flow, enhanced by
multiphase support, to simulate metal pad rolling in a three layer model
of a liquid metal battery.

\section{Numerical model}
The single-phase model developed to describe the magnetohydrodynamic
flow in one electrode of a liquid metal battery
 \cite{Weber2013} is enhanced by multiphase support. It
is implemented in the open source CFD library OpenFOAM
and based on the standard solver multiphaseInterFoam \cite{Maric2014}.

The fluid flow in a viscous, electrically conducting and incompressible
fluid of several phases is described by the Navier-Stokes equation
\begin{equation}\label{nse}
\frac{\partial(\rho \bi u)}{\partial t} + \nabla\cdot\left(\rho\bi u\bi u\right) 
 = - \nabla p +\nabla\cdot\left(\rho\nu\left(\nabla\bi u + (\nabla\bi u)^\intercal\right)\right) 
+\rho\bi g + \bi F_\text{L} + \bi F_\text{st}
\end{equation}
with $\bi u$, $t$, $\rho$, $p$, $\nu$,  $\bi g$,  $\bi
F_\text{L}$ and $\bi F_\text{st}$ meaning fluid velocity, time,
density, pressure, kinematic viscosity, gravity, Lorentz force
and surface tension force. The volume-of-fluid method is used 
to model the different phases of the battery. The phase fraction $\alpha_i$ denotes the
fraction of fluid $i$ in a single cell -- all variable fluid properties ($\rho, \nu, \sigma$)
can be defined by it. The evolution of the phase fraction $\alpha_i$ is calculated
by solving the equation
\begin{equation}
\frac{\partial\alpha_i}{\partial t} +\nabla\cdot(\bi u\alpha_i) = 0.
\end{equation}
The surface tension is computed using the continuum surface force
(CSF) model by Brackbill \cite{Brackbill1992}. It is modelled as a
volume force around the interfaces as 
\begin{equation}
\bi F_\text{st} = -\sum_i\sum_{j\neq i}\gamma_{ij}\kappa_{ij}\delta_{ij}
\end{equation}
with $\gamma_{ij}$ denoting the interface tension between
phases $i$ and $j$ (which is assumed to be constant) and $\kappa_{ij}$ 
the curvature of the interface. The force is applied locally around
the interface, using some kind of Dirac delta function
$\delta_{ij} = \alpha_j\nabla\alpha_i - \alpha_i\nabla\alpha_j$.
For a detailed description of
the implementation of the volume of fluid method and the surface
tension, see \cite{Ubbink1997,Rusche2002}.

The electric potential $\phi$ in all fluids is determined by solving the Poisson equation
\begin{equation}\label{mp:poisson}
\nabla\cdot(\sigma\nabla\phi)=\nabla\cdot(\sigma\bi u\times\bi B)
\end{equation}
with $\sigma$ and $\bi B$ meaning the electric conductivity and
magnetic field, respectively. Please note that Neumann boundary conditions force all
induced currents to close in the fluid, while Dirichlet boundary
conditions allow them to close in the solid current collectors. The
electric current $\bi J$ in the battery is then 
\begin{equation}\label{mp:ohm}
\bi J = -\sigma\nabla\phi + \sigma(\bi u\times \bi B).
\end{equation}
Using Green's identities we can express the magnetic vector potential $\bi A$ as
\begin{equation}\label{biotA}
\bi A(\bi r) = \frac{\mu_0}{4\pi}\int dV\frac{\bi J(\bi r')}{|\bi r -\bi
  r'|}
\end{equation}
and thus the magnetic field as $\bi B=\nabla\times\bi A + \bi B_z$. \frage{The 
vertical magnetic background field $\bi B_z$ is a constant. 
Equations
\ref{mp:poisson}, \ref{mp:ohm} and \ref{biotA} are solved for each time step
in order to compute finally the Lorentz force as $\bi F_\text{L} = \bi J\times 
\bi B$ as source term for the Navier-Stokes equation (equation \ref{nse})}.

\section{Instability mechanism}
\begin{figure}[b!]
\centering
\includegraphics[height=3.6cm]{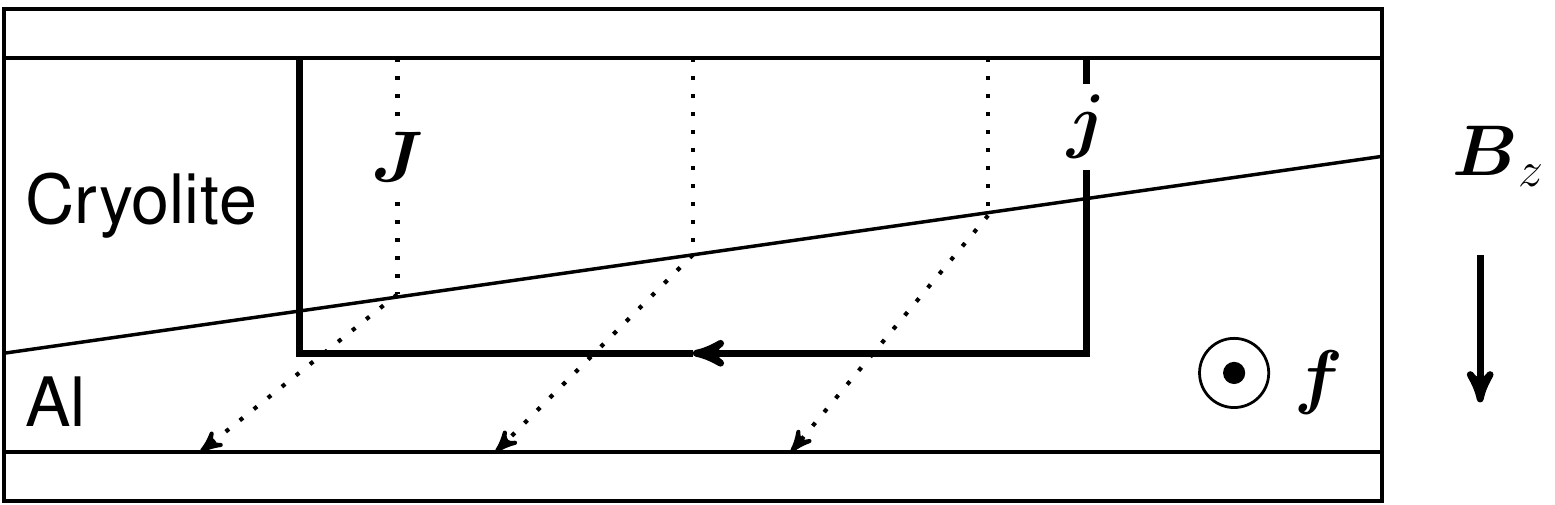}
\caption{Scheme of an aluminium reduction cell with inclined
  interface, global current $\bi J$, compensation                                                                                                                                 
current $\bi j$, magnetic background field $\bi B_z$ and Lorentz force
$\bi f$.}
\label{f:alCell}
\end{figure}
\begin{figure}[t!]
\centering
\includegraphics[height=6cm]{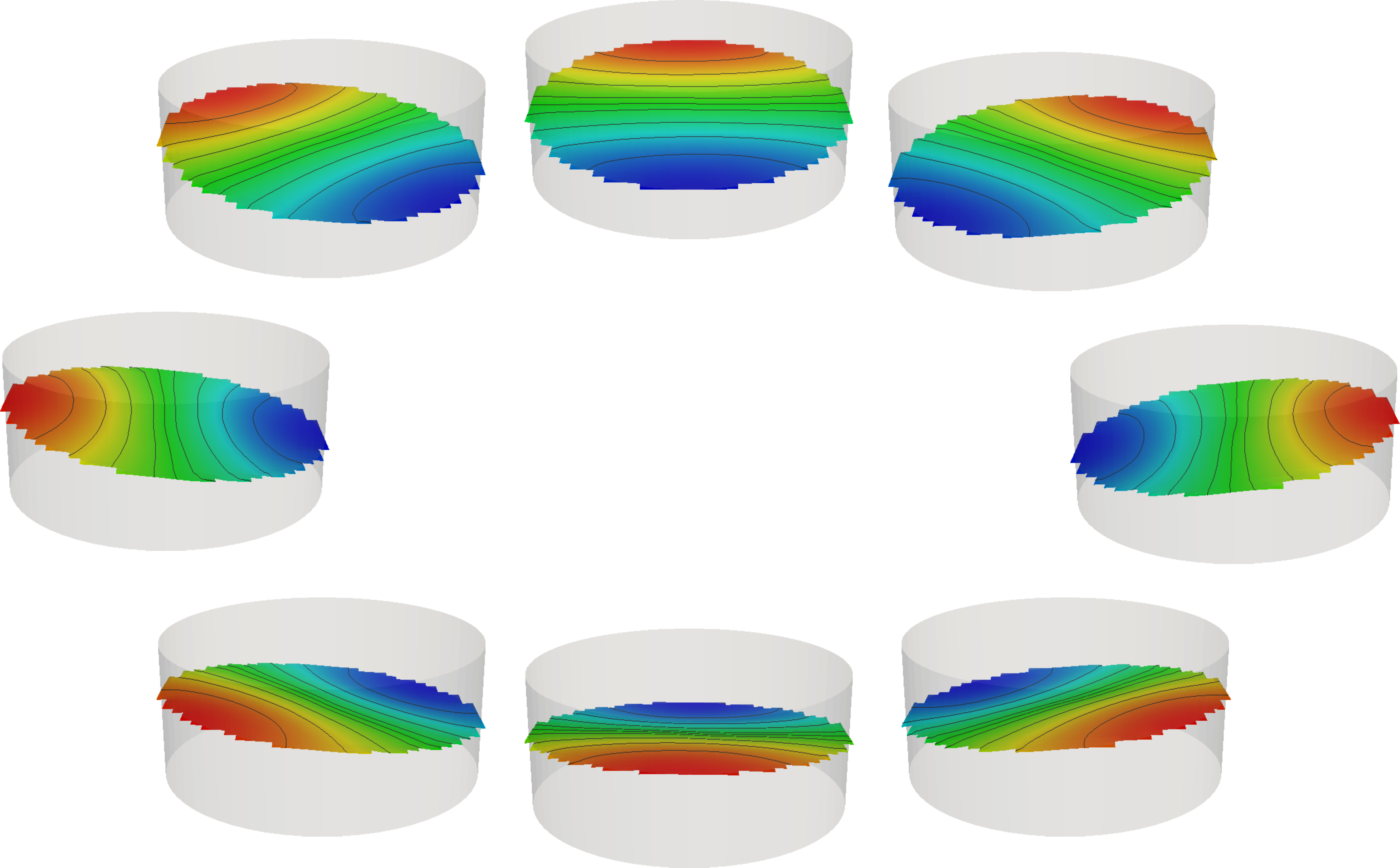}
\caption{Rotating inclined interface of a two layer system of an
  aluminium reduction cell.}
\label{f:rotate}
\end{figure}

We will firstly give a conceptual explanation of the instability mechanism with a simple model of an aluminium reduction
cell according to Sele \cite{Sele1977}. Such a cell consists of a liquid aluminium layer at the bottom
and a cryolite layer at the top (Fig. \ref{f:alCell}). A vertical
electrolysis current
is applied by carbon current collectors; the upper electrode is the
positive one. The electric conductivity of aluminium is $10^4$
times higher than that of cryolite. At certain conditions, a
deformation of the Al-cryolite
interface starts to rotate (Fig. \ref{f:rotate}); the initial
disturbance may even grow. This instability is known as metal pad
rolling, sloshing or long wave instability.

Sele \cite{Sele1977} was the first to provide a simple, understandable
model for the origin of this movement. For a flat
interface, the cell current $\bi J$ is purely vertical. An inclined interface leads
to a compensation current $\bi j$ comprising a strong horizontal fraction (Fig. \ref{f:alCell}).
If a
vertical magnetic background field $\bi B_z$ is present, the horizontal currents may interact
with it, leading to a Lorentz force pointing out of the plane (Fig. \ref{f:alCell}).
If we consider the Lorentz force to displace only the crest of
aluminium, it will generate a rotating wave. Note that the disturbed
current does not close in the cryolite due to its poor conductivity --
it has to close in the upper carbon current collector.

\begin{figure}[t!]
\centering
 \subfigure[]{\includegraphics[height=0.35\textwidth]{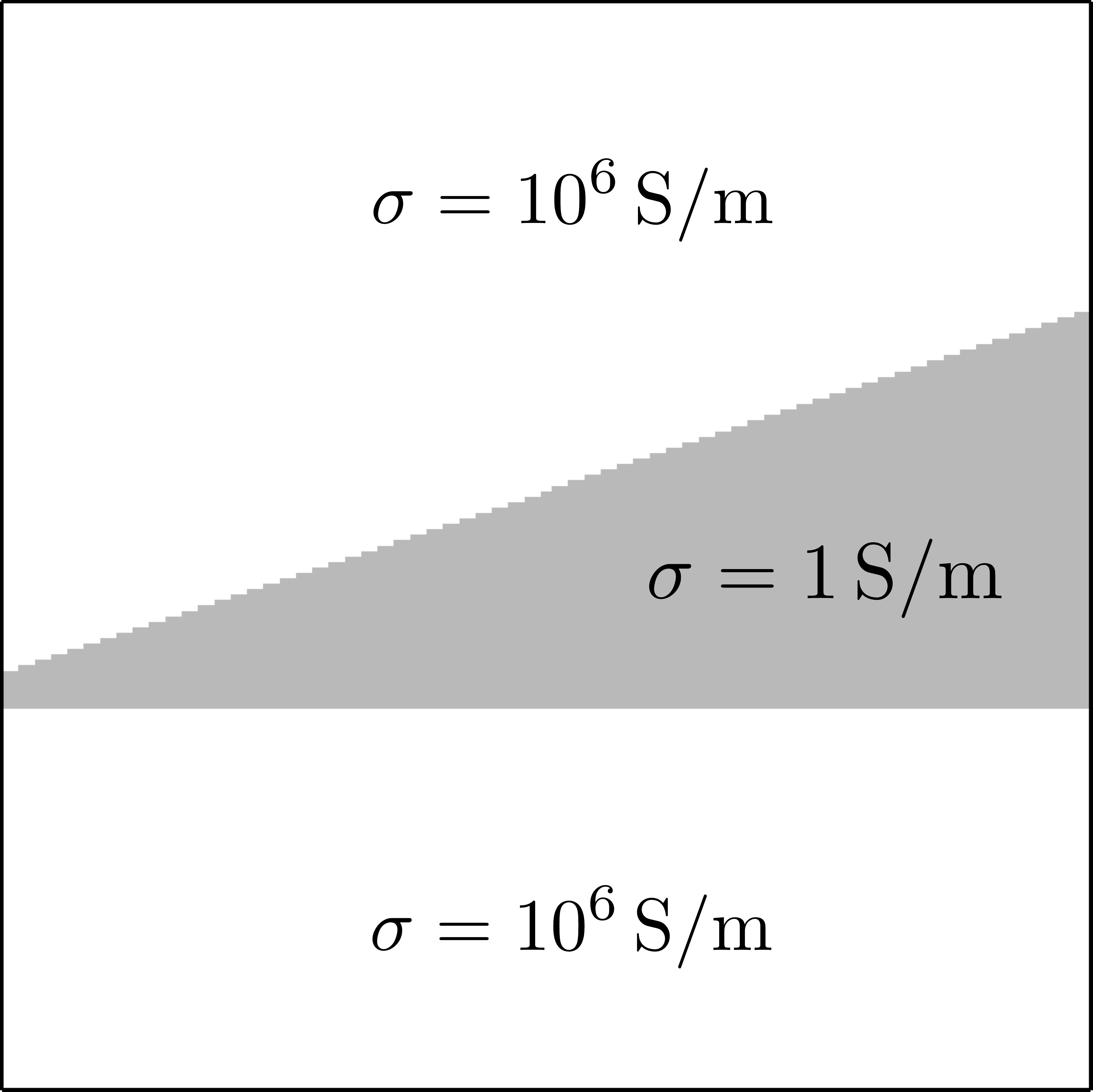}}\hfill
 \subfigure[]{\includegraphics[height=0.35\textwidth]{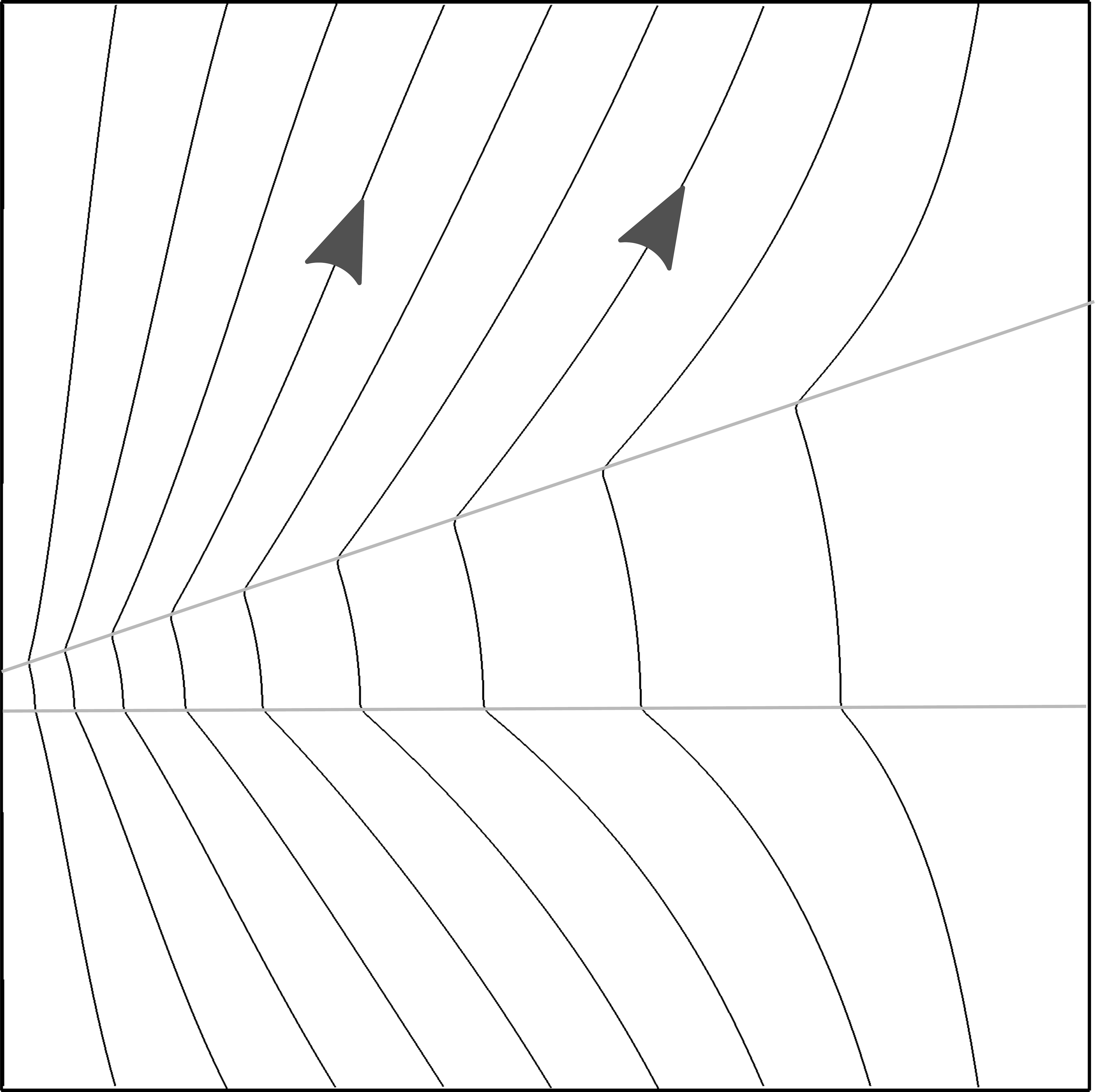}}\hfill
 \subfigure[]{\includegraphics[height=0.35\textwidth]{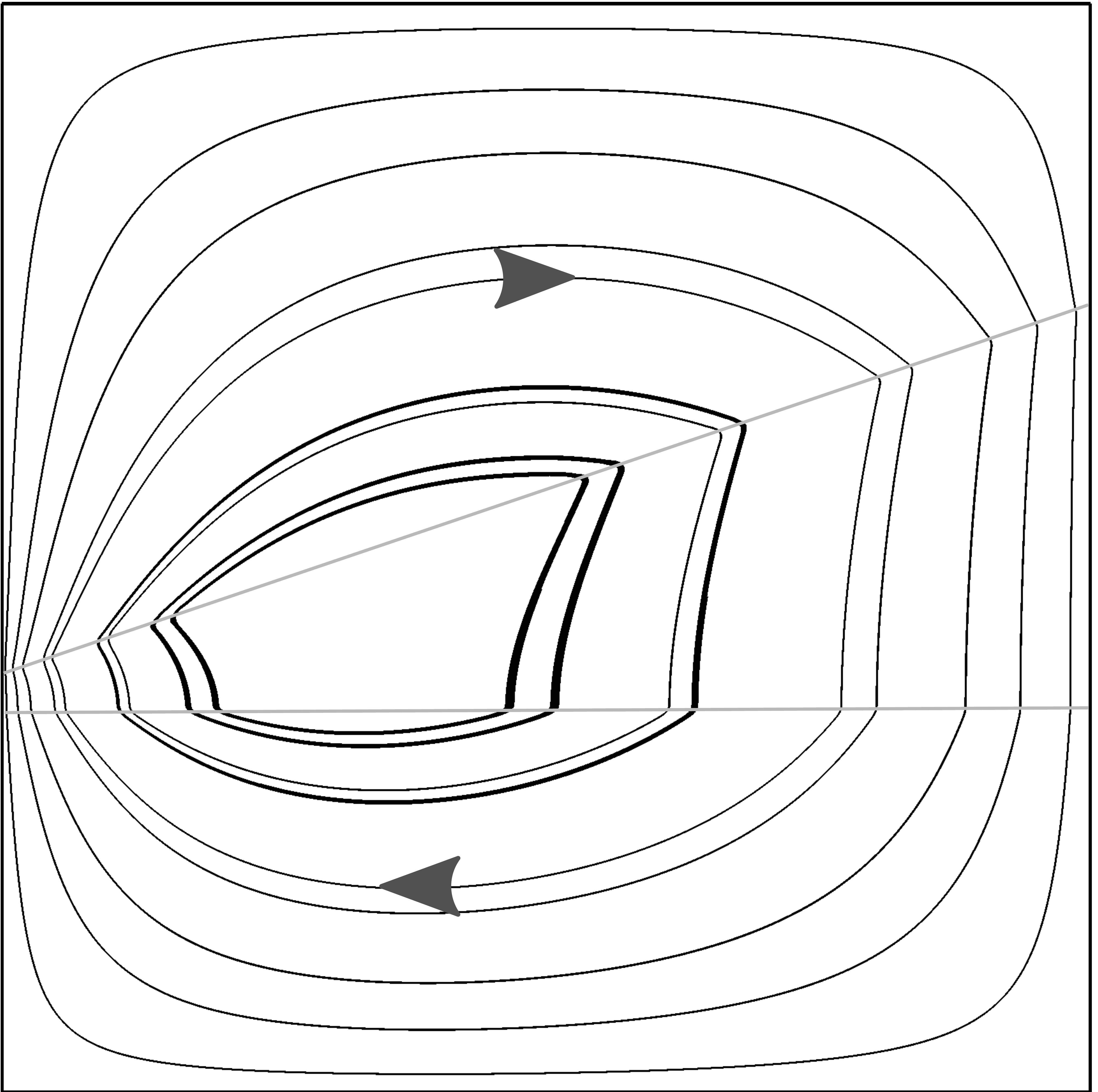}}\hfill
 \subfigure[]{\includegraphics[height=0.35\textwidth]{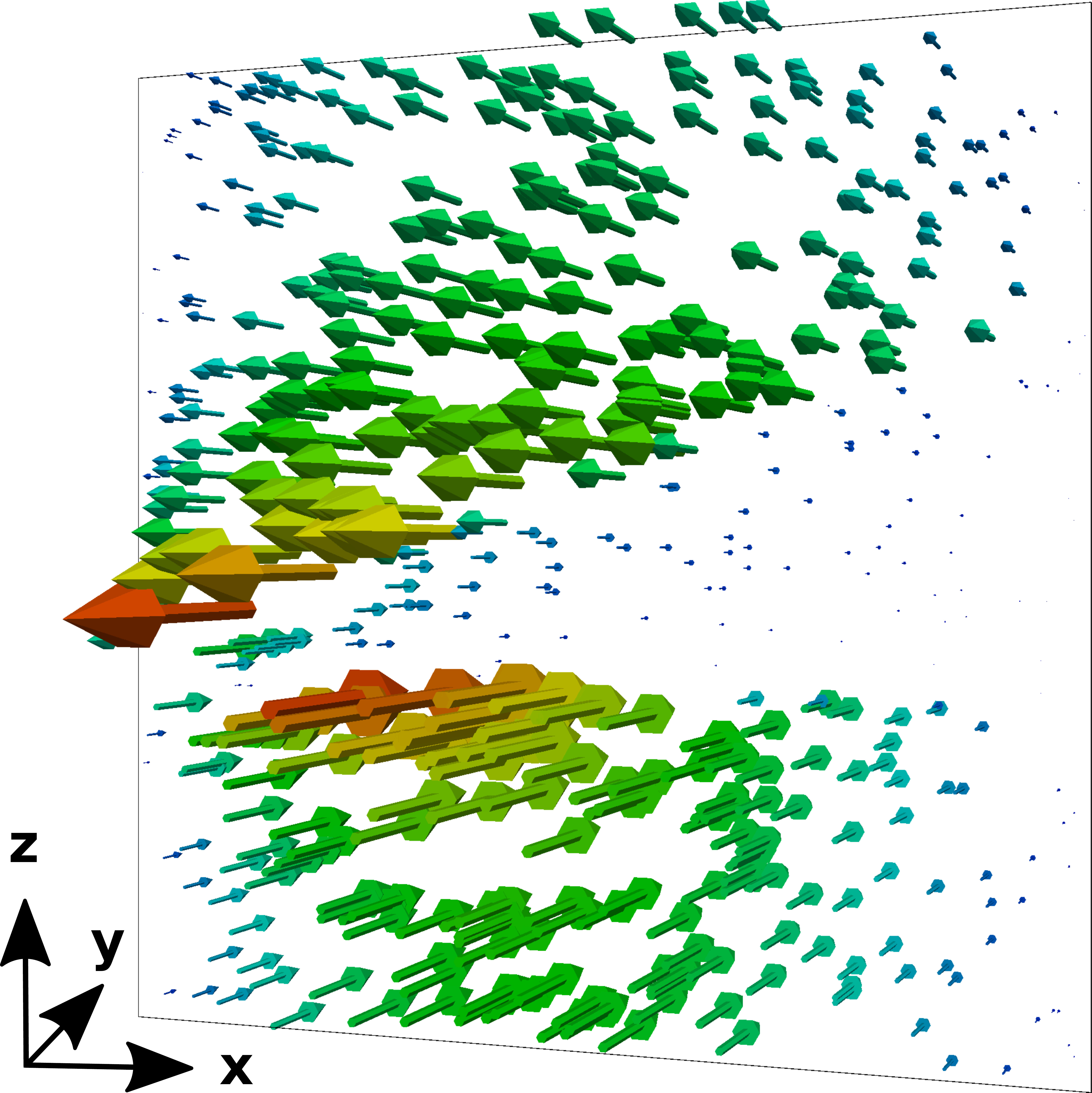}}
\caption{Conductivity (a), complete current (b) and compensation-current
(c) as well as Lorentz force of compensation-current and vertical field (d).
  The prescribed electrical current flows upwards, $\bi B_z$ is pointing upwards.}
\label{f:2dcurrent}
\end{figure}

In an LMB, the mechanism is quite similar. We consider a stratification
of three phases with inclined interface in 2D. The salt layer is poorly
conducting, compared to the metals (Fig. \ref{f:2dcurrent}a); \frage{the chosen
values do not belong to a certain LMB, but are of a typical order of magnitude.}
 For computing the electric potential
we use Neumann boundary conditions, i.e. the compensation current $\bi j$ has to
close in the liquid metals. This is quite realistic for tall electrodes; for
flat cells with bad conducting current collectors, a different choice of boundary conditions
is probably better suited. Fig. \ref{f:2dcurrent}b shows the full cell current $\bi J$
-- it is concentrated in the well conducting corner. Note that horizontal currents
appear only in the metals; in the salt layer the current is almost
vertical.
Fig. \ref{f:2dcurrent}c illustrates the perturbed
current $\bi j$.
Computing the cross product of a vertical magnetic background field
and the disturbed
current, we obtain the Lorentz force that is aligned with the
y-axis (Fig. \ref{f:2dcurrent}d). It
drives
the upper metal anti-clockwise and the lower clockwise. The crest
turns
then anti-clockwise -- it is pushed by the Lorentz force in the upper metal.

The Sele model presented above can describe how a rotating flow and also
a rotating wave develops. However, it does not explain why the wave amplitude
increases in time. This is assumed to be caused by a reflection of the wave crest 
at the solid walls \cite{Lukyanov2001, Molokov2011}. Using a wave equation 
approach \cite{Urata1985}, Bojarevics and Romerio extended the Sele model
by the influence of the cells aspect ratio. Due to the coupling of two
waves, cylindrical and quadratic cells are assumed to be always unstable 
\cite{Bojarevics1994} (neglecting several dissipation mechanisms).

\section{Metal pad rolling in a 3-layer system}
\begin{figure}[t!]
\centering
 \subfigure[]{\includegraphics[height=0.45\textwidth]{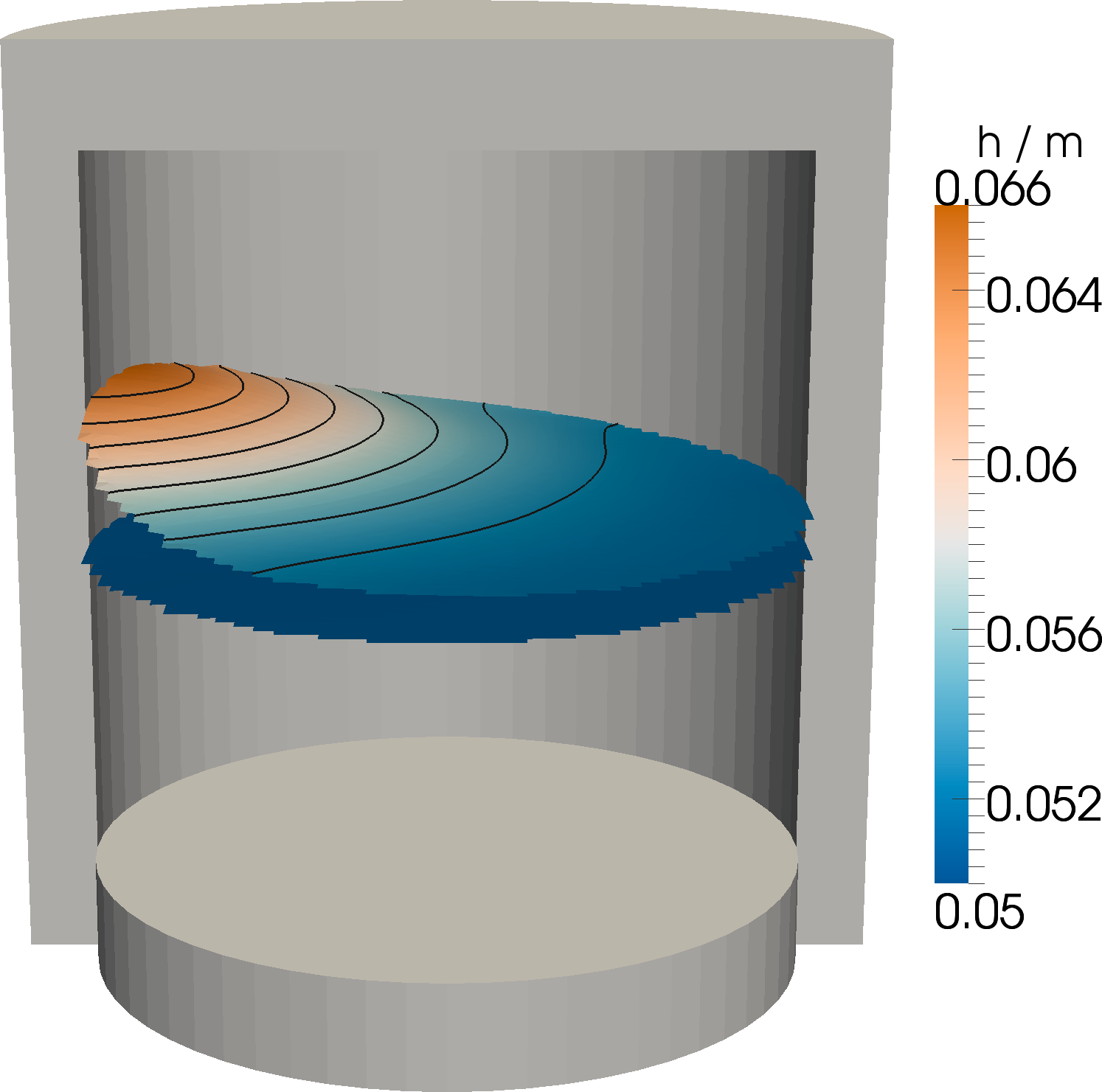}}\hfill
 \subfigure[]{\includegraphics[height=0.4\textwidth]{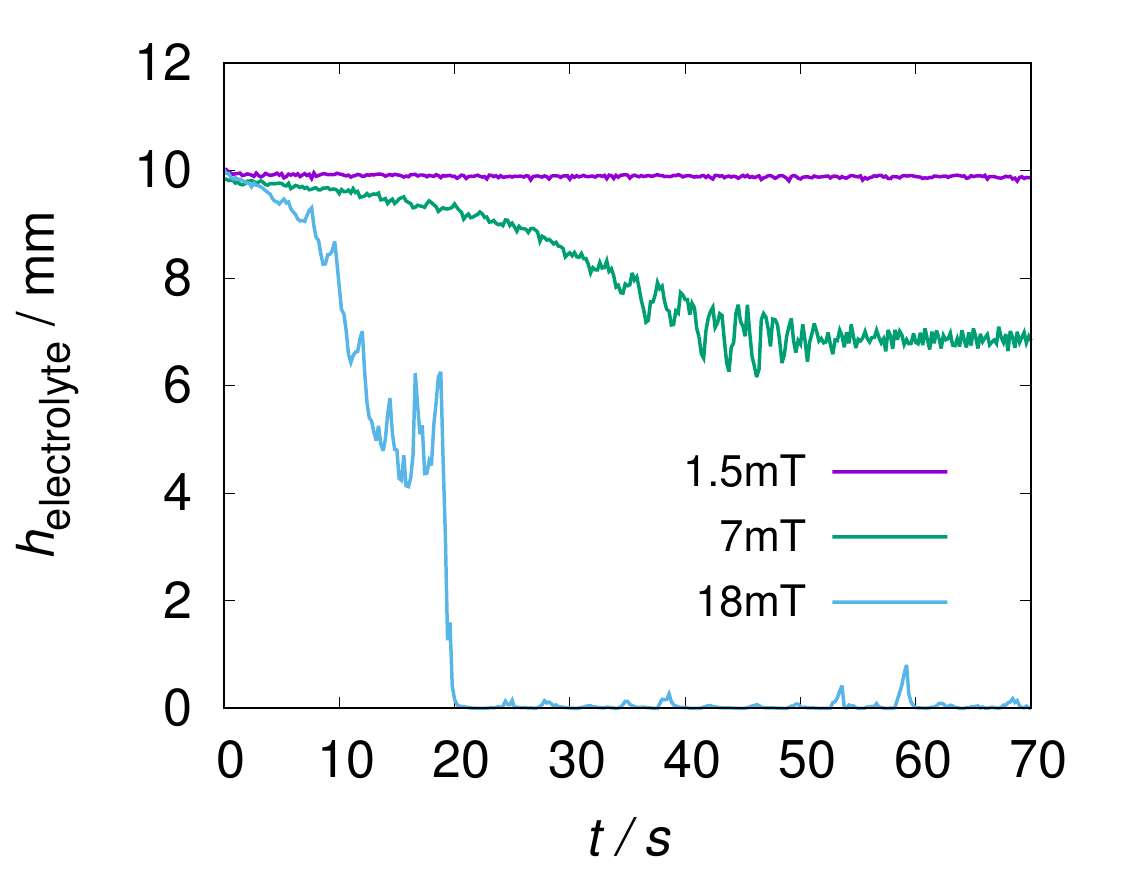}}\hfill
\caption{Illustration of the deformed interface due to metal pad
rolling for $B_z=10$\,mT (a) \frage{and minimal height of the electrolyte layer for
different magnetic background fields and $J=1$\,A/cm$^2$ (b). The
minimal distance between the two interfaces shown in (b) rotates; for
the rotation period see figure \ref{f:rotation}.}}
\label{f:height}
\end{figure}

First simulations of the 3-phase system of a Mg$|$NaCl-KCl-MgCl$_2$$|$Sb battery show
metal pad rolling -- see Fig. \ref{f:height}a. We use here a
simple cell model of height and diameter $h=d=10$\,cm, with a 1\,cm thick
electrolyte layer and 4.5\,cm thick electrodes. Denoting the upper metal by 1, the electrolyte by 2
and the lower metal by 3, the densities are \cite{Todreas2008,Kim2013b,Gale2004}
\begin{equation*}
  \rho_1 = 1577\, \text{kg/m}^3\qquad
  \rho_2 = 1715\, \text{kg/m}^3\qquad
  \rho_3 = 6270\, \text{kg/m}^3,
\end{equation*}
the viscosities
\begin{equation*}
\nu_1 = 6.7\cdot 10^{-7}\, \text{m}^2/\text{s}\qquad
\nu_2 = 6.8\cdot 10^{-7}\, \text{m}^2/\text{s}\qquad
\nu_3 = 2.0\cdot 10^{-7}\, \text{m}^2/\text{s}
\end{equation*}
and the electric conductivities
\begin{equation*} 
\sigma_1 = 3.6 \cdot 10^6\, \text{S/m}\qquad
\sigma_2 = 80\,  \text{S/m}\qquad
\sigma_3 = 8.7 \cdot 10^5\, \text{S/m}.
\end{equation*}
The mutual interface tension between the two phases $i$ and $j$ can be
estimated using the surface tensions of phase $i$ and $j$ as \cite{Israelachvili2011}
\begin{equation}
\gamma_{i|j} = \gamma_i + \gamma_j - 2.0\sqrt{\gamma_i\gamma_j},
\end{equation}
i.e.
\begin{equation*}
\gamma_{1|2} = 0.19\,\text{N/m}\qquad 
\gamma_{1|3} = 0.016\,\text{N/m}\qquad
\gamma_{2|3} = 0.095\,\text{N/m}.
\end{equation*}
\frage{We start the simulation with purely horizontal interfaces;
the initial disturbance is generated by small numerical errors.}
We apply a current of 78.5\,A which amounts to a current density of
about 1A/cm$^2$, that is typical for LMBs. 
In Fig. \ref{f:height}b
we show the minimal electrolyte layer thickness for a vertical magnetic
background field between 1.5 and 18\,mT. At a field of 7\,mT we observe
already a deformation of the interface; for values larger than 18\,mT
a short circuit appears. The values of the magnetic fields are chosen here
rather high in order to evidence clearly the effect of metal pad
rolling. Realistic vertical fields will probably be smaller than 10\,mT.

\begin{figure}[t!]
\centering
 \subfigure[]{\includegraphics[height=0.38\textwidth]{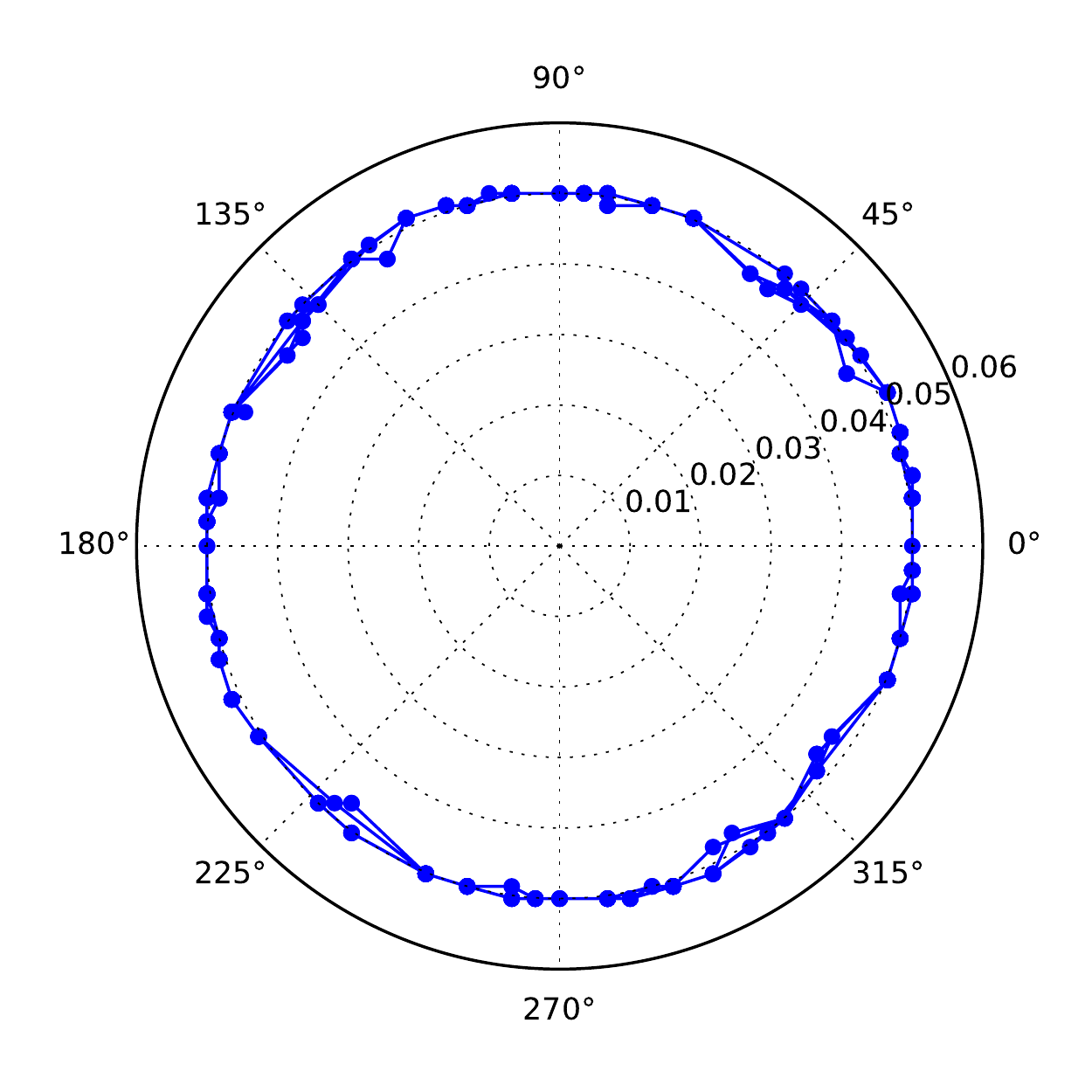}}\hfill
 \subfigure[]{\includegraphics[height=0.38\textwidth]{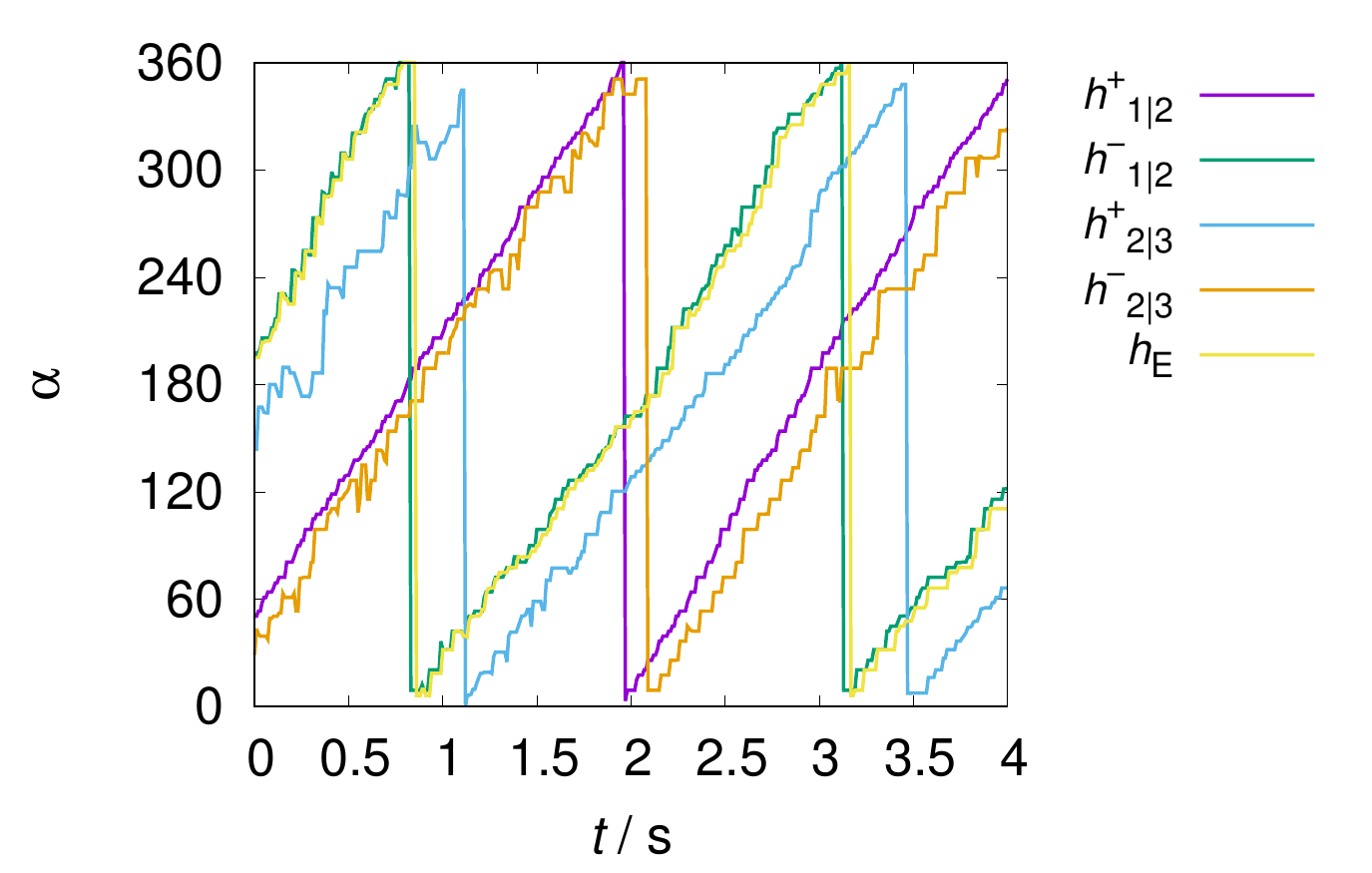}}\hfill
\caption{Radius of minimal electrolyte layer thickness (a) and angle of
smallest height $h_E$ as well as minimal $h^-$ and maximal $h^+$ elevation of both
interfaces for a simulation of $I=200$\,A (b).}
\label{f:rotation}
\end{figure}
The crest of electrolyte is well pronounced (Fig. \ref{f:height}a), the
trough however, is rather vast and flat. Both rotate with the same
speed and with a phaseshift of about 180$^\circ$. The minimal electrolyte
layer thickness is always located at the cylinder wall
(Fig. \ref{f:rotation}a). The same holds for the maximal elevation.

Compared to the upper interface, the lower one deforms only very
little. The maximum and minimum elevations all rotate with the same
speed, see Fig. \ref{f:rotation}b. Please note also that the maximum
of the lower interface ($h^+_{2|3}$) is always located under the minimum of the upper
interface ($h^-_{1|2}$). This promotes the short-circuit.

\begin{figure}[t!]
\centering
 \subfigure[]{\includegraphics[height=0.33\textwidth]{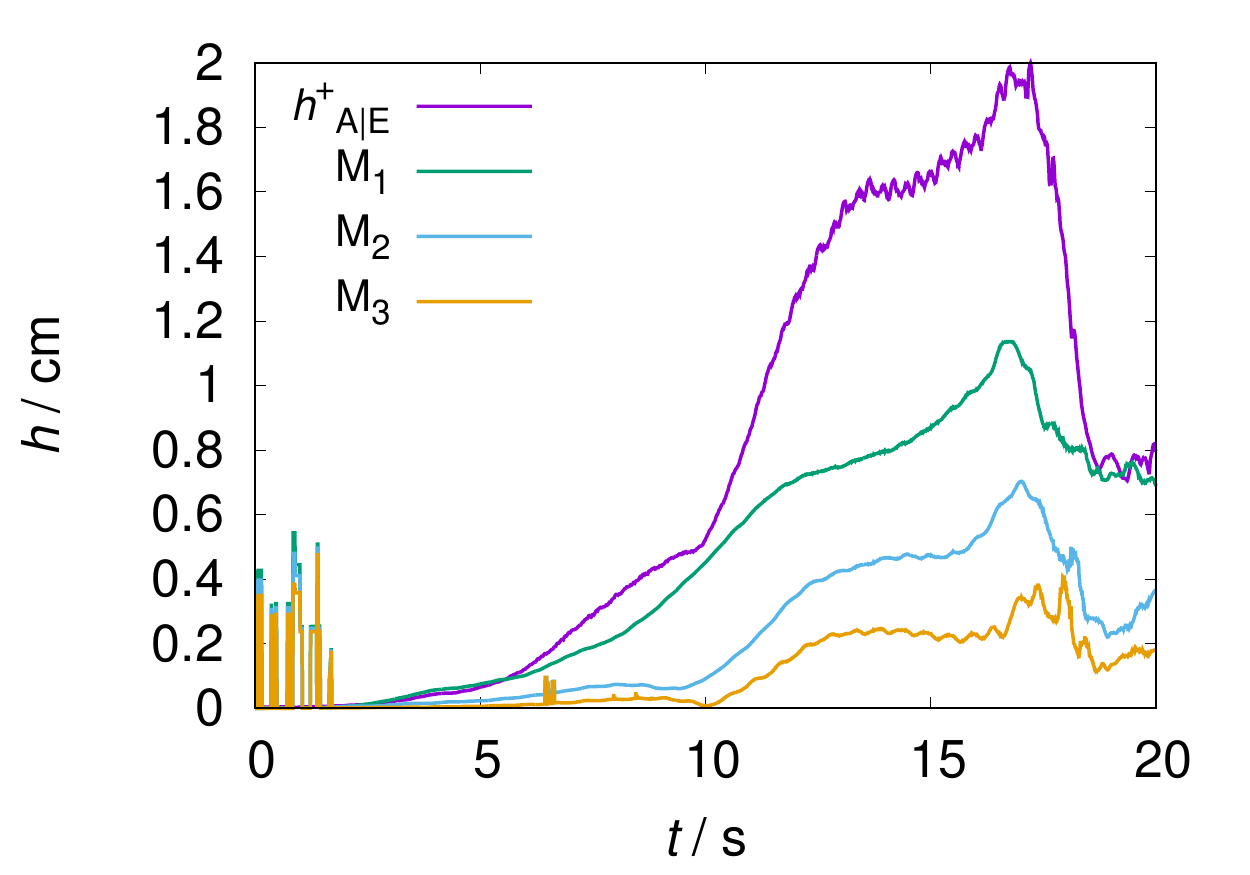}}\hfill
 \subfigure[]{\includegraphics[height=0.33\textwidth]{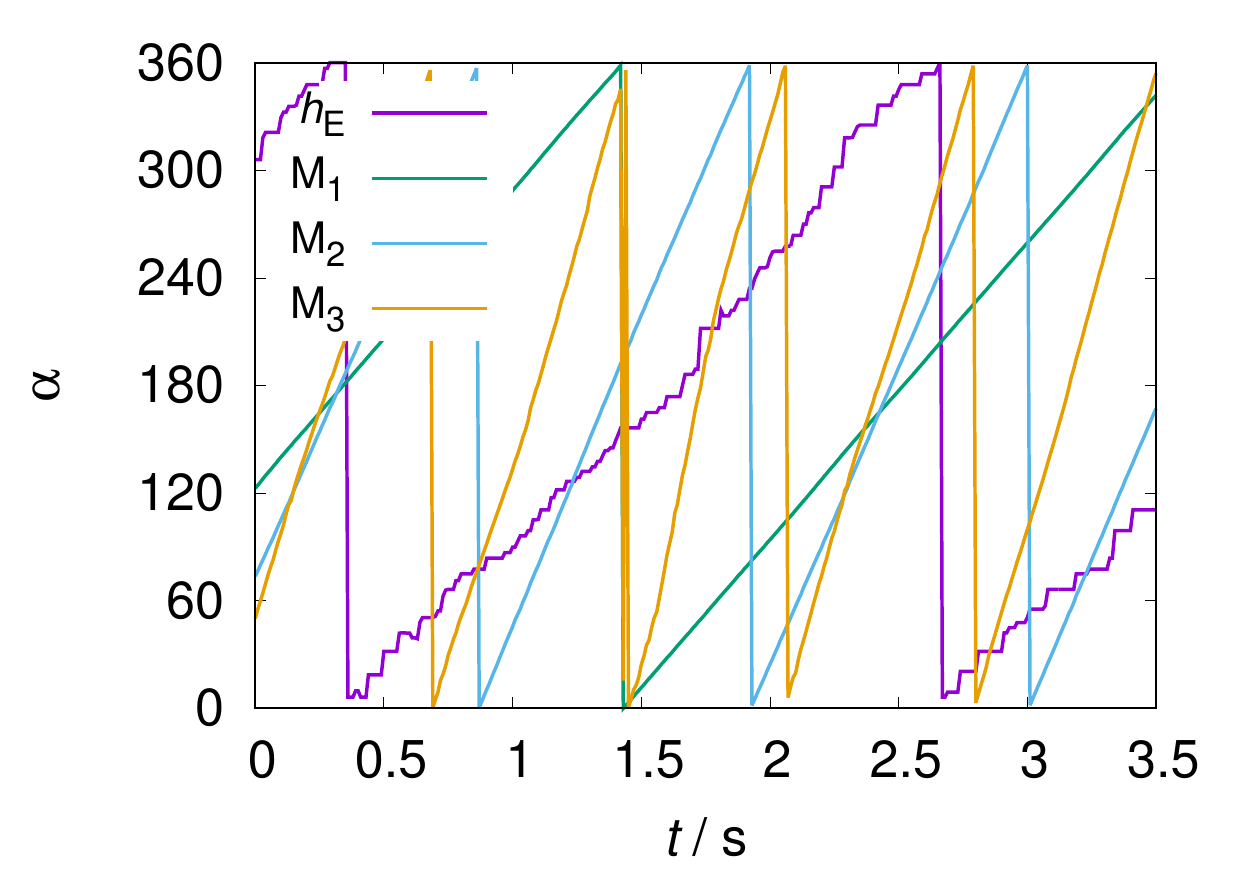}}\hfill
\caption{Amplitude (a) and angle of minimal electrolyte layer thickness and of the
  azimuthal modes one to three (b) for a simulation of $I=200$\,A.}
\label{f:modes}
\end{figure}

In order to study the exact shape of the upper interface, we 
decompose it into several azimuthal modes according to 
\begin{equation}
h_{1|2} = \sum_{m=0}^\infty (a_m\sin(m\alpha + \beta_{m}))
\end{equation}
with $\alpha$ denoting the angle. At a first glance, the interface 
deformation in Fig. \ref{f:height} seems to have only one crest and 
trough. This corresponds to the mode $m=1$ with $a_m$ 
denoting its amplitude and $\beta_{m}$ its phase shift. The mode
$m=2$ has then two crests and troughs, and so on. We illustrate now
in Fig. \ref{f:modes}a the amplitude of the first three modes. As 
expected, the highest amplitude is observed for $m=1$; however, the
modes $m=2$ and $m=3$ still reach about 50 and 25\,\% of the first one.
The higher modes rotate with a shorter period (Fig. \ref{f:modes}b):
the simulated periods for the first three modes are: $T_1
=2.18$\,s, $T_2 = 1.08$\,s and $T_3 = 0.76$\,s with $I=200$\,A and 
$B_z=10$\,mT.

It is known, that the metal pad roll in aluminium reduction cells oscillates
approximately with the frequency of gravity waves. Since the bottom 
fluid in LMBs is heavy, the $2|3$ interface deforms very little. This allows 
to estimate the frequency of the rotating wave with
\begin{equation}\label{eqn:omega}
\omega^2 =
\dfrac{(\rho_2-\rho_1)gk}{\frac{\displaystyle\rho_1}{\displaystyle\tanh(kh_1)}+\frac{\displaystyle\rho_2}{\displaystyle\tanh(kh_2)}},
\end{equation}
that is recognized as the dispersion relation of gravity waves in 
a 2 layer system \cite{Cappanera2015}. With wavenumber $k=36.8$\,m$^{-1}$ 
as the first zero of the derivative of the Bessel-function $J_1'(kd/2) = 0$
 and $g=9.81$\,m/s$^2$, we find $T_1 = 2.28$\,s. This confirms that 
the mode $m = 1$ indeed is a gravity wave. The periods $T_2 \approx T_1/2$
 and $T_3 \approx T_1/3$ for $m=2$ and $m=3$ are typical for nonlinear 
harmonics of the fundamental $m=1$ wave.
\begin{figure}[t!]
\centering
\includegraphics[height=0.4\textwidth]{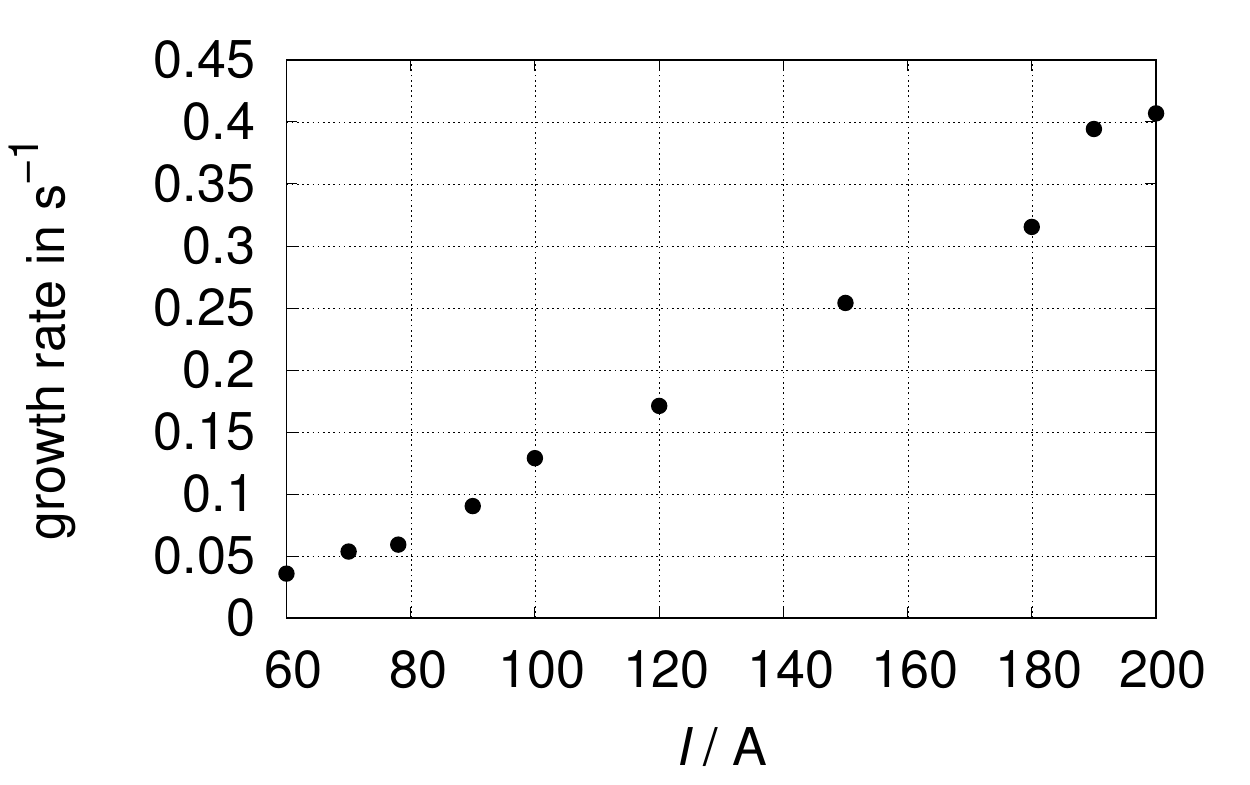}
\caption{Growth rate of the electrolyte deformation for different
  electrical currents and $B_z=10$\,mT.}
\label{f:growthrate}
\end{figure}

In a last step we study the growth rate of the deformation of
the interface. Often it is difficult to obtain this growth rate due
to the limited electrolyte thickness and possible oscillations of the
interface (Fig. \ref{f:height}b). Fig. \ref{f:growthrate} shows the
growth rates for different cell currents. The relation between both
appears to be linear; however, the result should be treated with caution,
because of the mentioned difficulties in fitting the values.

\section{Conclusions and outlook}
We have developed an integro-differential equation model to simulate 
metal pad rolling in liquid metal batteries. First simulations are 
very promising: the expected long wave instability appears (only) at
the upper interface of the three-phase system. This is not surprising, 
since the lower metal remains rather stable due to its high density.
As a result, we explained the origin of metal pad rolling in LMBs in
close analogy to the known mechanism in aluminium reduction cells. We
further described the interface deformation, the growth rate of the
instability as well as the period of rotation. Our
simulations are another step forward to understand the complex fluid
mechanics involved in the operation of large liquid metal batteries.

It is intended to investigate metal pad rolling in liquid metal
batteries further. Especially the influence of battery current,
magnetic background field and height of the electrolyte layer will be
analysed. Comparative studies with the spectral code SFEMaNS
\cite{Guermond2009} are planned for the near future. A validation of
our code with experimental findings is in prospect in the long run.

\Thanks{This work was supported by Helmholtz-Gemeinschaft Deut\-scher
  Forschungs\-zentren  in frame of the Helmholtz Alliance ``Liquid
  metal technologies'' (LIMTECH). The computations were performed on
  the HPC-Cluster ``Taurus'' at the Center for Information Services
  and High Performance Computing (ZIH) at TU Dresden and on the
  cluster ``Hydra'' at Helmholtz-Zentrum Dresden -- Rossendorf. We
  gratefully acknowledge fruitful discussions with Valdis Bojarevics,
  Andreas Bund, Lo{\"i}c Cappanera, Julien Commenge, Jochen
  Fr\"ohlich, Douglas Kelley, Cornel Lalau, Steffen Landgraf, Michael
  Nimtz, Marco Starace, J\=anis Priede and Oleg Zikanov on several
  aspects of metal pad rolling and liquid metal batteries and Henrik
  Schulz for the HPC support.}

\clearpage
\bibliographystyle{mhd}
\bibliography{literatur}

\lastpageno	

\end{document}